\documentclass[
floatfix,
aps,
pra,
amsmath,
twocolumn,
]{revtex4-1}

\usepackage{graphicx}
\usepackage{bm}
\usepackage{amsmath}
\usepackage{amssymb}
\usepackage{array}
\usepackage{bm}

\usepackage{stackengine}

\usepackage{xcolor}

\usepackage [autostyle, english = american]{csquotes}
\MakeOuterQuote{"}

\newcommand{\be}{\begin{equation}}
\newcommand{\ee}{\end{equation}}

\newcommand{\ket}[1]{|{#1}\rangle}
\newcommand{\bra}[1]{\langle {#1} |}
\newcommand{\braa}[1]{\langle {#1} }

\usepackage[normalem]{ulem}

\newcommand{\bea}{\begin{eqnarray}}
\newcommand{\eea}{\end{eqnarray}}

\def\rrr#1\\{\par
\medskip\hbox{\vbox{\parindent=2em\hsize=6.12in
\hangindent=4em\hangafter=1#1}}}

\begin{document}

\title{Steering random spin systems to speed up the quantum adiabatic algorithm}

\author{A. Bar\i\c{s} \"Ozg\"uler}
\affiliation{Department of Physics, University of Wisconsin--Madison, Madison, WI 53706}
\author{Robert Joynt}
\affiliation{Department of Physics, University of Wisconsin--Madison, Madison, WI 53706}
\author{Maxim G. Vavilov}
\affiliation{Department of Physics, University of Wisconsin--Madison, Madison, WI 53706}


\begin{abstract} 
A general time-dependent quantum system can be driven fast from its initial ground state to its final ground state without generating transitions by adding a steering term to the Hamiltonian.  We show how this technique can be modified to improve on the standard quantum adiabatic algorithm by making a single-particle and cluster approximation to the steering term.  The method is applied to a one-dimensional Ising model in a random field.  For the limit of strong disorder, the correction terms significantly enhance the probability for the whole system to remain in the ground state for the proposed non-stoquastic  annealing protocol.  We demonstrate that even when transitions occur for stronger interaction between qubits, the most probable quantum state is one of the lower energy states of the final Hamiltonian.  Since the method can be applied to any model, and more sophisticated approximations to the steering term are possible, the new technique opens up an avenue for the improvement of the quantum adiabatic algorithm.   
\end{abstract}

\date{November 8, 2018}

\maketitle

\section{Introduction}

There is a furious race underway to construct the first practical quantum computer.  To complement this, there is a large research effort to broaden the class of problems that can be attacked by these machines.  A very promising direction is optimization problems.  One of the leading candidate methods for solving such problems on a quantum computer is the quantum adiabatic algorithm (QAA) \cite{farhi2001quantum}, in which the ground state of a simple quantum system is slowly transformed into the solution of the optimization problem.  There have been extensive studies of the QAA on classical computers \cite{mcgeoch2014adiabatic} and open-system  quantum annealing devices intended to solve similar problems have been  constructed \cite{johnson2011quantum,boixo2013experimental,boixo2014evidence}. The QAA exploits the adiabatic theorem and uses the fact that the ground state of appropriate quantum Hamiltonians correspond to difficult classical optimization problems, for which the standard classical search algorithms are inefficient due to the complicated landscape for the cost function \cite{tanaka2017quantum,lucas2014ising}.  The difficulty in demonstrating the QAA is the presence of small energy gaps that can lead to generalized Landau-Zener-Stueckelberg-Majorana (LZSM) tunneling \cite{Landau1932,Zener1932,Stueckelberg1932,Majorana1932}. Once the tunneling occurs, the system leaves the instantaneous ground state, probably for good, and the algorithm breaks down.

In spin models, we may look more closely at the degrees of freedom that produce the dangerous avoided crossings.  The classic LZSM problem can be thought of as a single spin-1/2 particle in a time-dependent magnetic field that reverses the spin direction.  This is the local single-particle case.  In the other limit, we may imagine a crossing of two levels whose energies are very close, but whose spatial configurations differ by the rearrangement of many spins, perhaps well-separated in space.  This is the non-local case.  Both contribute to unwanted tunneling. 

In this paper, we propose a modification of the QAA that largely eliminates local LZSM tunneling.  This modification requires accurate control of individual qubits that was demonstrated recently in various systems, including trapped ions \cite{Vittorini2014_pra}, Rydberg atoms \cite{Maller2015_pra} and superconducting qubits \cite{Chow2009_prl}.  In the conventional annealing protocol, the system is prepared in a strong field along the $x-$direction without interaction, then the field is slowly changed to the final field and the interaction is turned on.  During this process, a time-dependent gauge term causes transitions between the instantaneous eigenstates of the Hamiltonian.  This term is proportional to the Berry curvature \cite{berry1984quantal,Avron2010,Avron2011,Gritsev2011,Xu2014} and its effect was recently 
investigated in superconducting devices with a single qubit \cite{Schroer2014} and interacting qubits \cite{Roushan2014}.  We demonstrate that with the proper compensation of this topological term, qubits acquire protection against excitation processes, increasing the probability for the system to remain in the ground state even for short annealing times.
This approach may also point the way toward more general improvements of quantum adiabatic algorithms.  

\section{Method}

The Hamiltonian in our approach is defined on the time interval $0 \leq t \leq t_a$, where $t_a$ is the annealing time and it has the form:
\begin{equation}
H_{qaa}(t/t_a) =  f_i(t/t_a)H_{i}+f_f(t/t_a) H_{f}+H_{s}(t),
\label{eq:eq1}
\end{equation}
Here $H_i$ and $H_f$ are time-independent Hamiltonians that represent a simple problem and a difficult optimization problem, respectively.  The scalar functions $f_i$ and $f_f$ satisfy the boundary conditions: $f_i(0)=f_f(1)=1$ and $f_i(1)=f_f(0)=0$.  However, we adjust these functions rather than choosing the customary linear-in-time forms. $H_s$ is the steering term and key to our approach.  The idea of adding an additional term to the Hamiltonian is not new and has been used to convert a stoquastic Hamiltonian to a non-stoquastic Hamiltonian \cite{hormozi2017nonstoquastic,vinci2017non}, while modifications to the annealing schedule have been used to add quantum fluctuations \cite{kadowaki1998quantum}. It has also been used in the method of shortcuts to adiabaticity and quantum critical points \cite{del2012Assisted,Takahashi2013,del2013shortcuts,del2014tracking,del2014universality,del2015controlling,rams2016inhomogeneous,Kazutaka2017,sels2017minimizing}. Our method is to make a local approximation to the exact formula for the counterdiabatic driving Hamiltonian, defined in the following paragraph.

We construct  $H_s$ using a result from adiabatic population transfer theory and counterdiabatic driving \cite{demirplak2003adiabatic, berry2009transitionless}. If a time-dependent Hamiltonian $H_0$ has instantaneous eigenstates $|n(t)\rangle$ such that $H_0(t)|n(t)\rangle=E_n(t)|n(t)\rangle$, then we can define the steering Hamiltonian as
\begin{equation}
H_1(t) = i \hbar \sum_{m = 2}^{2^L} \frac{\ket{m}\bra{m}\partial_t H_0 \ket{1}\bra{1}}{E_1 - E_m}  + \text{(h.c.)}.
\label{eq:eq2}
\end{equation}
The modified Hamiltonian $H(t) = H_0(t)+H_1(t)$ drives the ground state $|1 \rangle$ of $H_0$ without any transitions.  If the initial state at $t = 0$ is the ground state of $H_0$, then the solution of the time-dependent Schr\"odinger equation at $t_a$ is the ground state of $H_0$.  We could take $H_0 = f_i(t/t_a)H_{i}+f_f(t/t_a) H_{f}$ and $H_s = H_1$, and this would yield the solution of the optimization problem with certainty, but unfortunately the computation of $H_1$ is not efficient. Instead, we propose a local approximation to $H_1$. We note that for single spin-1/2 particle at site $k$ with Hamiltonian  
$H_0^{(k)}(t) = \bm{B^{(k)}}(t)\cdot{\bm{\sigma^{(k)}}}/2$ the steering term is
\begin{equation}  
H_{0,s}^{(k)}(t)=  \frac{1}{2 (B^{(k)}(t))^2} (\bm{B^{(k)}}(t)\times \partial_t \bm{B^{(k)}}(t)) \cdot\bm{\sigma^{(k)}}
\label{eq:eq3}
\end{equation}
and we may correct for an arbitrary random magnetic field on an array of spins by summing over $k$.

To illustrate our method we choose the one-dimensional random-field Ising model (RFIM) on a ring of $L$ spins: 
\begin{equation}  
H_{f} =  \sum_{k=1}^{L} h_k \sigma_z^{(k)} +  J \sum_{k=1}^{L}  \sigma_z^{(k)} \sigma_z^{(k+1)}
\label{eq:eq4}
\end{equation}
with periodic boundary conditions understood.  The $h_k$ are chosen uniformly from the interval $[-1,1]$. The width of the disorder distribution sets the energy scale.   
The initial Hamiltonian is chosen as usual to be a uniform transverse magnetic field
\begin{equation} 
H_{i} =  h_0 \sum_{k=1}^{L} \sigma_x^{(k)}.
\label{eq:eq5}
\end{equation}
In the calculations below we take $h_0=10$.

The RFIM at $J=0$ has the simple solution $\braa{\sigma_z^{k}}\rangle = -h_k/|h_k|$, while the $J \rightarrow \infty$ limit is an antiferromagnet.  At small $J$, $J << h_{av}$ ($h_{av}$, average random field, $\sim$ 1/2 in this paper), the ground state has just a few spins that deviate from the $J=0$ solution at sites $k$ where $|h_k|$ happens to be small. The spin at site $k$ feels a time-dependent effective field with a $z$-component given by the sum of $h_k$ and $J [\braa{\sigma_z^{(k-1)}(t)}\rangle + \braa{\sigma_z^{(k+1)}(t)}\rangle]$, where $\braa{\sigma_z^{(k\pm 1)}(t)}\rangle$ are the time-dependent expectation values of the $z$-components of the neighboring spins.  When the magnitude of the total effective field (including the $x$-component) becomes small, the gap becomes small and the QAA can fail.  This is the type of failure that our local approximation for $H_1$ should be able to fix.  At larger $J$ values, ($J$ of order 1)  there will be larger clusters of spins that deviate from the $J=0$ solution.  This will create situations where there are small energy gaps separating states that differ by many spin flips.  Our single-spin approximation for the steering term is then not expected to work, and more sophisticated approximations are required.  We will later present a cluster method that is a step in this direction.     

It is clear that the steering method is applicable in principle to any model that includes a random field.  Our choice of the RFIM is motivated by the facts that it has a relatively small number of parameters, is simple to simulate numerically, and the statistical properties of the final Hamiltonian of Eq.~\eqref{eq:eq4} have been well studied. By the standards of the field, the one-dimensional RFIM is fairly simple but it has nevertheless served as a common testbed for the QAA. 

Notice that $H_i$ and $H_f$ are both stoquastic \cite{bravyi2017complexity} but the introduction of $H_s$ makes the Hamiltonian non-stoquastic.  This is somewhat similar to a previous study, \cite{hormozi2017nonstoquastic}, but our motivation for introducing the additional term is quite different.

We choose $f_i(t) = \cos^2(\pi \tau /2)$ and $f_f(t) = \sin^2(\pi \tau /2)$, where $\tau \equiv t/t_a$. The initial behavior of $f_f$ and the final behavior of $f_i$ are quadratic; this is chosen so that $H_s(t=0) = H_s(t=t_a) = 0$ and the derivatives provide slow start and stop.  These choices, together with Eq.~\eqref{eq:eq3}, give 
\begin{equation}
H_s(\tau)=\sum_{k=1}^{L}\frac{-h_0 h_k \pi \sin(\pi \, \tau )}{4 \, t_a[h_0^2\cos^4(\pi \, \tau /2 )+h_k^2\sin^4(\pi \, \tau /2)]} \\ \, \sigma_y^{(k)}.
\label{eq:eq6}
\end{equation}
Since $t_a$ can be small, the size of the steering term can be large.  Of course an arbitrarily large $H_s$ is unphysical. Ultimately, the interesting parameter range for the QAA is when $t_a$ is large.  In this case the steering term is typically small compared to the other terms in the Hamiltonian.

\section{Results}
With these definitions we solve the time dependent Schr\"odinger equation for $H_{qaa}$ numerically \cite{johansson2012qutip,JOHANSSON20131234}.  For comparison purposes it is useful to solve the same instance of the problem with the above definition of $H_s$ ("with steering")
and setting $H_s=0$ ("without steering").  We also define the success probability, i.e., the probability to be in the ground state at the end of the evolution, as $P_1=|\braa{1}\ket{\psi(t=t_a)}|^2$.

\begin{figure}[!htbp]
\includegraphics[width=0.49\textwidth]{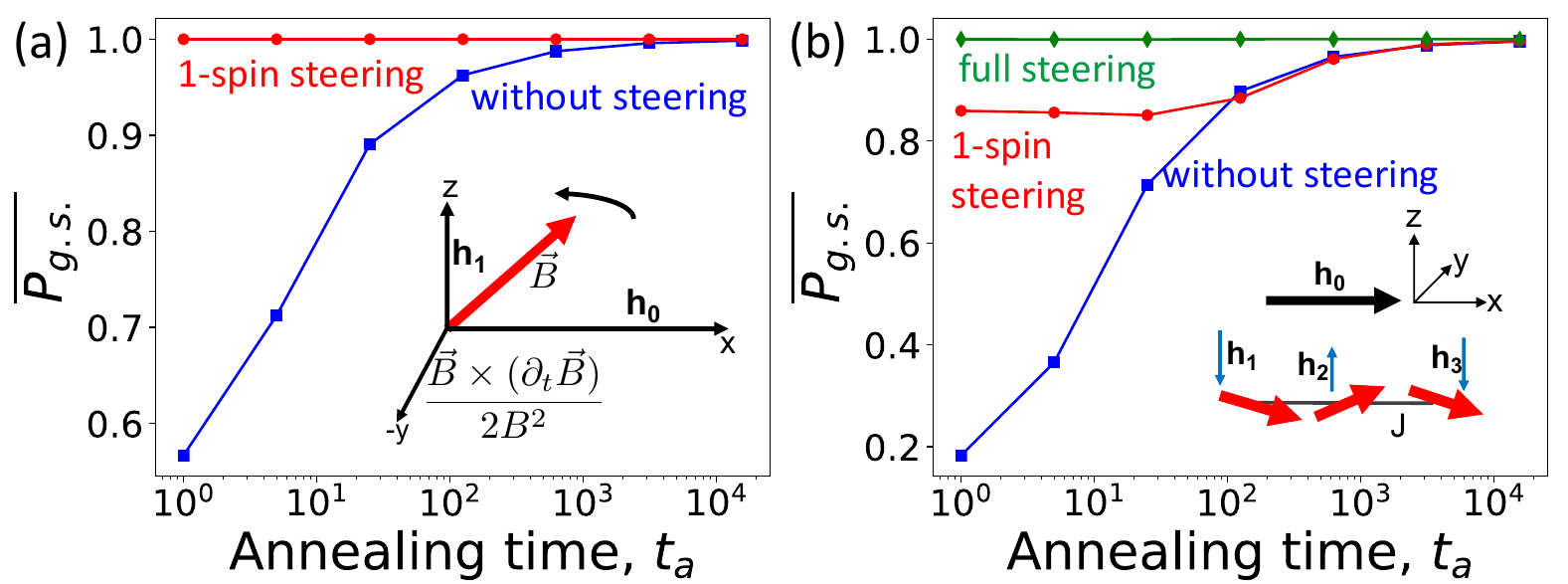}
\caption{\label{fig:fig1} (Color online) Average ground state probability as a function of the annealing time $t_a$. The $h_k$ are chosen uniformly from the interval $[-W,W]$, where $W = 1$. All energy variables are measured in units of $W$, and time variables are measured in units of $\hbar/W$ throughout the paper. (a) $L = 1$. In the inset, the red magnetic field vector rotates from $x$ to $z$ direction in the standard quantum annealing process. The steering field applied in the $-y$ direction suppresses transitions to the excited states. (b) $L = 3$, $J = 0.1$. The green diamond curve is the result of the application of Eq.~\eqref{eq:eq2}, the exact Berry formula.  The inset shows the sketch of the open chain of 3 spins considered here.}
\end{figure}

\subsection{Small Systems}
In Fig.~\ref{fig:fig1} we report results for the systems with $L = 1$ and $L = 3$ using Eq.~\eqref{eq:eq2}.  In part (a) of the figure, we show the fundamental effect of steering. The system finds the ground state independent of the annealing time to within our numerical accuracy for this case, which is to say 1 part in $10^9$. In part (b) of the figure, we compared the 1-spin steering with the case of no steering applied and with the "full steering". Full steering is the exact application of Eq.~\eqref{eq:eq2}. It is the basis of the cluster approach that we present in the later part of the paper. 

\subsection{Comparison to Other Methods}
\begin{figure}[!htbp]
\includegraphics[width=0.44\textwidth]{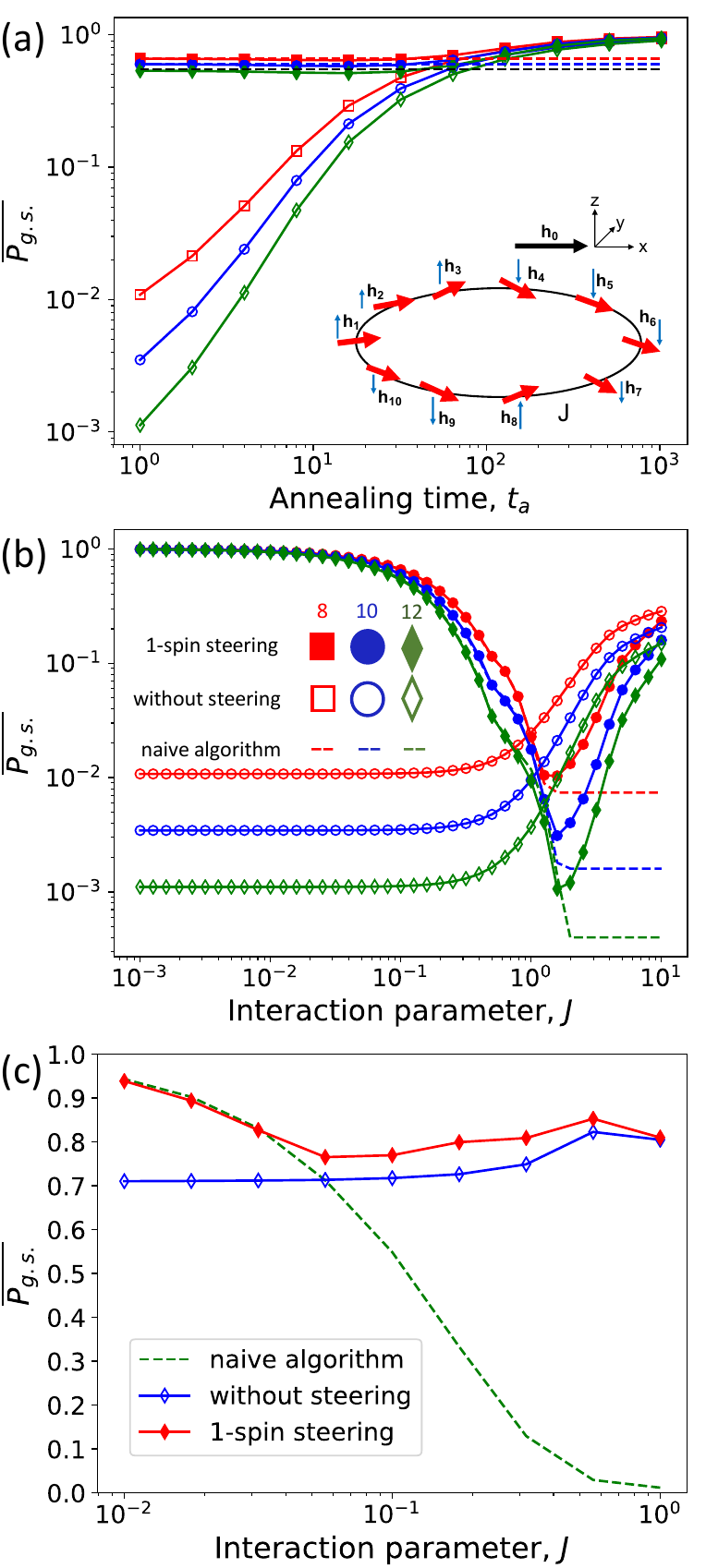}
\caption{\label{fig:fig2} (Color online) (a) Average ground state probability as a function of the annealing time $t_a$. $L$ = 8 (square), $L$ = 10 (circle), $L$ = 12 (diamond) compared for $J$ = 0.1. (b) Average ground state probability as a function of the interaction parameter $J$ for a short annealing time $t_a = 1$. The red (upper), blue (middle) and green (lower) dashed lines show the naive algorithm results for $L = 8, 10, 12$, respectively. (c) Average ground state probability as a function of the interaction parameter $J$ for a longer annealing time $t_a = 100$.}
\end{figure}

Small systems are only of interest for illustration purposes.  Practical applications require larger systems.  Because of the need to average over disorder realizations, we are limited to $L \leq 12$.  A sketch of the system we consider is shown in the inset of Fig.~\ref{fig:fig2}(a) for $L$ = 10. In Fig.~\ref{fig:fig2}(a), we present how the average ground state probability changes as a function of the annealing time for a weak interaction ($J = 0.1$). Especially for short annealing times, the probability of achieving the ground state and thereby successfully solving the optimization problem is quite small without steering. It is greatly enhanced by steering for short and long annealing times. In Fig.~\ref{fig:fig2}(a) we also show as dashed lines the result of a "naive" classical algorithm in which we choose the solution of the non-interacting system: $\braa{\sigma_z^{k}}\rangle = -h_k/|h_k|$. This solution is obtained by choosing $J = 0$ in the problem Hamiltonian $H_f$, Eq.~\eqref{eq:eq2}, and applying the steering, Eq.~\eqref{eq:eq6}. The steered QAA outperforms this algorithm in the range $t_a > 10^2$ for $J = 0.1$.

\begin{figure}[!htbp]
\includegraphics[width=0.44\textwidth]{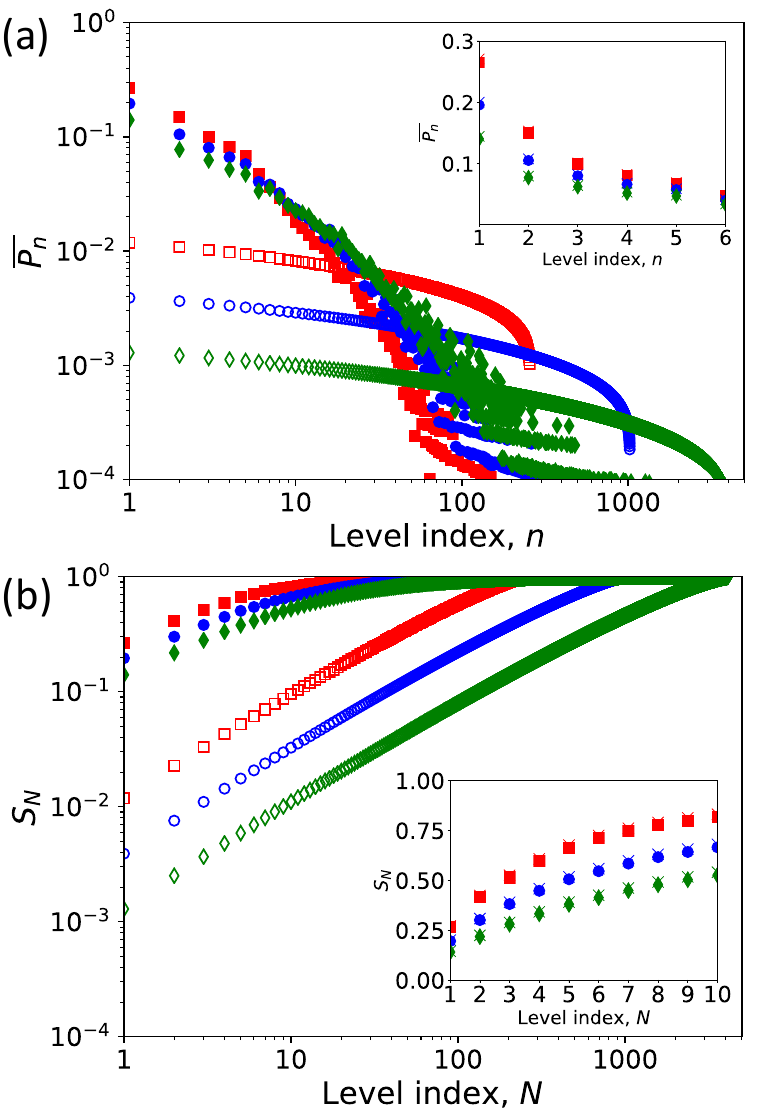}
\caption{\label{fig:fig3} (Color online) Same markers are used in this figure as in Fig.~\ref{fig:fig2}(a) and (b) for the standard QAA and the steered QAA. For the naive algorithm, red (upper), blue (middle) and green (lower) "x" markers are used in the insets for $L = 8, 10, 12$, respectively. In the insets, the naive algorithm is compared with the steered QAA. $t_a$ = 1, $J$ = 0.3. Several system sizes are shown.  (a) The probability distribution over all final eigenstates $|n(t_a)\rangle$ as a function of the level index $n$, computed by comparing the results of the QAA with an exact calculation. $P_n=|\langle \psi(t_a)|n(t_a)\rangle|^2$. The effect of steering is to squeeze the width of the probability distribution by two orders of magnitude and in the direction of the ground state. (b) Cumulative probability distribution. $S_N = \sum\limits_{n = 1}^N \overline{P_n}$. With the steered algorithm, the chance to find one of the low-lying states is significantly enhanced.} 
\end{figure}

\begin{figure*}[!htbp]
\includegraphics[width=0.99\textwidth]{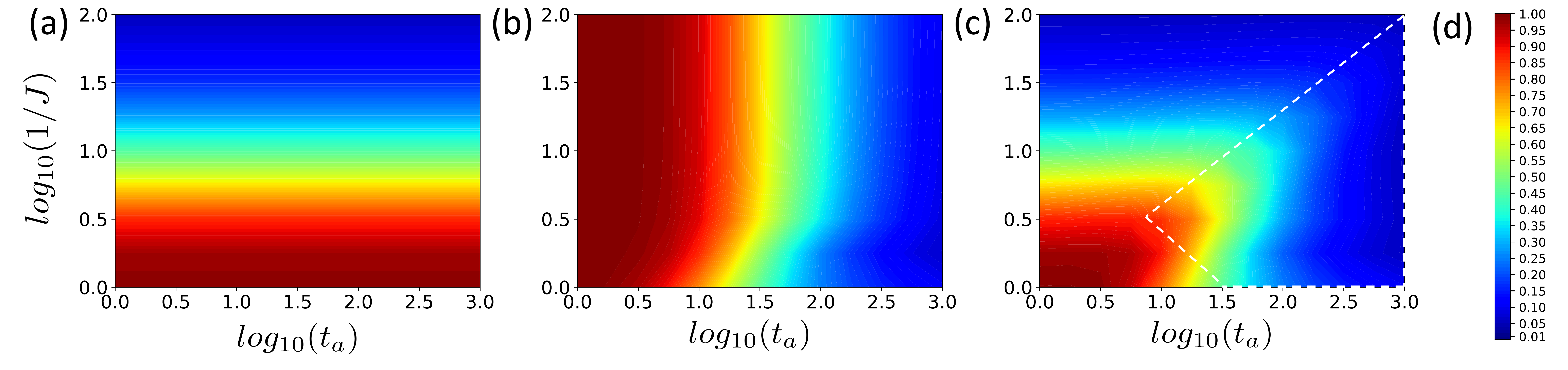}
\caption{\label{fig:contour_plot} (Color online) Average infidelity as a function of $1/J$ and $t_a$. $L = 12$. Plots for (a) the naive algorithm, (b) the standard QAA, and (c) the steered QAA. In the region covered by the white dashed lines, the steered QAA gives higher fidelity than the other two algorithms. (d) The colorbar shows the infidelity values.} 
\end{figure*}
 
When the interaction becomes stronger, the low-lying states have a more entangled character; they cannot be written, even approximately, as product states. Thus the local steering algorithm becomes ineffective. This is shown for a short annealing time $t_a = 1$ in Fig.~\ref{fig:fig2}(b), where the average ground state probability is plotted as a function of $J$.  We see a crossover at $J \sim 1$ from a regime in which steering is effective to a regime where it is not.  It is interesting that the addition of $H_s$ does not improve the QAA for $J \geq 2$, and can even degrade the performance.  We attribute this to the fact that the system, for part of its evolution, is trying to find the ground state of a Hamiltonian $H_i+H_f+H_s$ that is somewhat further from the problem Hamiltonian compared to $H_i+H_f$.  The "recovery" of the steered Hamiltonian at larger $J$ is presumably due to the ground state being a locally perturbed antiferromagnetic state, close once more to a product state.  For such a short annealing time, of course both the steered QAA and the standard QAA perform relatively poorly.  This can be seen by plotting the results for the naive algorithm, shown by the dashed lines.  Obviously, the results of this algorithm are independent of $t_a$.  Its success is similar to that of the steered QAA for $J < 1$.  For larger values of $J$, the naive algorithm performs poorly, as expected from the fact that it ignores interactions.

In Fig.~\ref{fig:fig2}(c) the annealing time is longer: $t_a = 100$. We see similar trends overall - steering becomes ineffective at larger $J$.  This plot does show clearly that there are definite differences between the standard QAA and the steered QAA at intermediate annealing times.  

\subsection{Distribution over Low-lying States}
Next, we consider how the introduction of a moderate interaction ($J=0.3$)  modifies the final distribution of the probability over all states both with and without steering, using a short annealing time $t_a=1$.  Recall that $L$ is the number of spins and the total number of levels is $2^L$, which is the size of the classical problem.  In Fig.~\ref{fig:fig3} we plot probabilities $P_n$ of all states, defined as 
$P_n=|\langle \psi(t_a)|n(t_a)\rangle|^2$, and the cumulative probability, defined as $S_N = \sum\limits_{n = 1}^N \overline{P_n}$.  The states $|n(t_a)\rangle$ are eigenstates of $H_f$ and they are arranged in order of increasing energy. $|\psi(t_a)\rangle$ is the final state computed in the QAA. This is done for several system sizes.  Of course to obtain these data we must also solve the problem exactly for $|n(t_a)\rangle$, so this limits the size of systems we can treat. Again we average over $10^4$ realizations of the disorder for each curve shown.

The effect of steering on the QAA is very dramatic.  Roughly speaking, for all system sizes the width of the probability distribution is squeezed down towards the ground state by two orders of magnitude by steering the QAA. The chance of making a serious error and ending in a state with high index is greatly reduced.  If we think of the system as diffusing from one instantaneous eigenstate to another during the course of a computation, it seems that the effect of steering is to reduce the diffusion rate regardless of whether the system is close to the ground state or not.  

Certain final states or groups of final states appear to be favored, and the groups are somewhat different for the steered and unsteered cases.  We can speculate that these states represent local energy minima.  The unsteered algorithm may in fact be superior in escaping local minima that come from extended eigenstates while the steered algorithm is more effective at avoiding local minima that come from more localized eigenstates. 

On the other hand, for these values of $t_a$ and $J$, the advantage of the steered QAA over the naive algorithm is marginal --- the data points nearly overlap.  In the next subsection we investigate when the results for these two algorithms separate.

\subsection{Regime of Superiority of Steered QAA over Other Methods}
Figs. 1, 2 and 3 demonstrate that steering can improve the QAA substantially for $J\leq 0.3$ and $t_A \leq 10$.  However, our results so far leave open the possibility that a combination of the standard QAA and the naive algorithm could give a roughly comparable performance to the steered QAA.  We now show that this is not the case. In Fig.~\ref{fig:contour_plot} we present contour plots of the infidelity for the naive algorithm, the unsteered QAA and the steered QAA as a function of the two key parameters $t_a$ and $J$. This allows us to locate the range in which the performance of the steered QAA is superior.  This is the interior of the dashed white region in Fig.~\ref{fig:contour_plot}(c).  Since this is a log-log plot, the range of parameters inside the region is quite large.

The key point is that steering is in fact effective when the spin interacts with its neighbors.  It becomes entangled with neighboring spins and its state can no longer represented by a pure state on the surface of the Bloch sphere, but one may still define an effective field.  When the magnitude of the total effective field is small, a small gap in the excitation spectrum is likely.  This is obviously the dangerous case.  Our results show that steering is also effective in this situation.  The steered QAA is superior to the unsteered QAA in all cases.  The improvement is particularly dramatic when $t_a$ is small, but even at moderate values the improvement is substantial.

\subsection{Cluster Steering}
\begin{figure}[!htbp]
\includegraphics[width=0.43\textwidth]{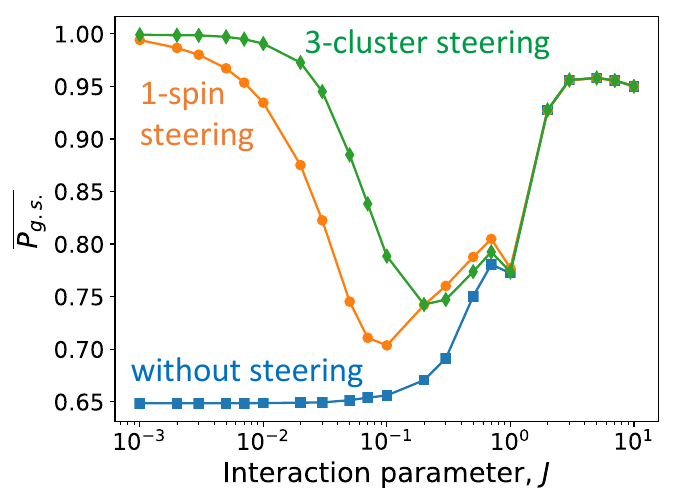}
\caption{\label{fig:fig4} (Color online) Average ground state probability as a function of the interaction parameter $J$ for the QAA without steering, with 1-spin steering, and with cluster steering.  Cluster steering improves the results for $J \leq 0.2$ $t_a = 128$, $L$ = 12.}
\end{figure}

One of the advantages of the steering method is that it is susceptible to systematic improvement.  The results presented so far are only those that follow from a single-particle approximation to the steering Hamiltonian.  In this subsection, we present our results for a more sophisticated approximation that we call cluster steering.  This is defined as follows.  The spin which has minimum random field (whose direction is therefore likely to be determined by the interaction) is identified. This spin and its two neighbors are considered as a cluster. The cluster steering term is found from Eq.~\eqref{eq:eq2}. In this approach, while the cluster steering is being applied to the spin trio, 1-spin steering is applied to each spin in the rest of the chain. There are 12 spins in the chain and $10^4$ realizations are performed. In Fig.~\ref{fig:fig4}, the two types of the steering are compared with the case of no steering. At small $J$, the curves with steering coincide and, at stronger $J$, all curves go up. The latter happens because in this regime the spectrum becomes more regular with level repulsion. However, the steering of weak clusters helps to maintain the system in the ground state even for intermediate strengths of interaction. With the cluster approach, the ground state probability does not drop to smaller values sharply. When $J$ is small, the ground state probability curve is more flat comparing to the curve of 1-spin steering.

\section{Conclusion}
We have demonstrated significant improvements in the QAA for random-field spin systems with relatively weak interactions.  This is done by adding a term to the Hamiltonian that suppresses transitions representing local spin re-orientations.  When the interactions become stronger, the low-energy eigenstates become more extended and the technique in the approximation used here becomes ineffective. In other words, the method is good for insulating phases and not for metallic phases of disordered systems.  However, the steering concept itself, as represented by the correction term in Eq.~\eqref{eq:eq3} is not at all limited to local modifications of the problem.  We have made a cluster expansion to construct a less local form of the operator in Eq.~\eqref{eq:eq6}. It should also be possible to work out ways of improving the steering so that it is effective in metallic phases as well.

We have not yet investigated systematically the crucial question of how the improvements in the algorithm scale with system size.  The local nature of the improvements of the steering would suggest at least a constant speedup comparing to the standard annealing procedure.  Of course in practical calculations even a constant speedup is very desirable, as long as the constant is big.  For certain problems, we have shown that two orders of magnitude can be achieved. 









The protocol is applied to a particular configuration of the final Hamiltonian, where both local fields and couplings between the spins are exactly determined by the corresponding classical optimization problem. To evaluate the performance of the algorithm for different problems with similar structure, we assume that the optimization problems represent an ensemble of random Hamiltonians.  The success is determined both by the quantum fidelity of the final state and by the fraction of successful solutions out of the ensemble. For $t_a$ fixed at 1, we have the following comparisons for the standard and steered QAAs. Out of exponentially large system size $2^L$ with $L=12$, we find with probability above 99\% that the system is in one of 21 low energy states when $J = 0.1$. For $J = 0.3$ and the same $L$, we find one of the 398 low energy states with probability above 99\% for the QAA with 1-spin steering. For the unsteered algorithm, the corresponding values are too large --- 3949 and 3929, respectively. For the QAA with 1-spin steering, the probability to find one of the lowest 1\% of $2^L$ (with $L=12$) energy states is 99.7\% when $J = 0.1$. When $J = 0.3$, the probability becomes 81\%. For the unsteered algorithm, the corresponding probabilities are only 3\% and 4\%, respectively.  Thus, by controlling $3L$ local fields, we are guaranteed to find one of the low energy states out of $2^L$ states.

We have also compared the steered QAA to a naive classical algorithm that works only for weak interactions.  Combining all our results shows that there is a substantial range of parameters for which the steered QAA outperforms both the standard QAA and the naive algorithm.

\section*{Acknowledgments} 
 
We are thankful to Sergey Knysh and Vadim Smelyanskiy for fruitful discussions. The simulations were performed using the computing resources and assistance of the UW-Madison Center For High Throughput Computing (CHTC). The work was supported by NSF EAGER Grant No. DMR-1743986.

\bibliographystyle{apsrev-nourl}

\bibliography{steering}

\end{document}